\documentclass{article}
\usepackage{RR}
\usepackage[latin1]{inputenc}
\usepackage{latexsym}
\usepackage[cmex10]{amsmath}
\usepackage{amstext,amsthm,amsfonts,amssymb}
\usepackage{pstricks,pst-node,pst-tree}
\usepackage[dvips]{epsfig}
\usepackage[dvips]{graphicx}
\usepackage{graphics}
\usepackage{epsf}
\usepackage{float}
\usepackage{color}
\usepackage{epic,eepic}
\usepackage{fancyhdr}
\usepackage[gen]{eurosym}
\usepackage{hyperref}
\usepackage{subfig}
\usepackage{placeins}
\usepackage{mathenv}
\newtheorem{thm}{Theorem}

\begin{document}
\RRdate{Octobre 2010}
\RRtitle{Probabilit\`e de Coupure pour Traitement Multi-Cellulaire Conjoint en Fading de Rayleigh} 

\RRetitle{Outage Probability for Multi-Cell Processing under Rayleigh Fading} 

\RRauthor{Virgile Garcia\thanks{virgile.garcia@inrialpes.fr}, Nikolai Lebedev\thanks{lebedev@cpe.fr}, Jean-Marie Gorce\thanks{jean-marie.gorce@insa-lyon.fr} } 

\authorhead{Garcia \& Lebedev \& Gorce}
\titlehead{Outage Probability for Multi-Cell Processing under Rayleigh Fading}

\RRtheme{\THCom} 
\RRprojet{SWING} 
\RRresume{Ce papier fourni une expression théorique de la probabilit\`e de coupure en capacit\`e pour le traitement multi-cellulaire conjoint (CoMP) en voie descendante et en conditions de fading de Rayleigh. Il est montr\`e que la formulation th\`eorique correspond parfaitement aux simulations. Dans un contexte où les communications sont limit\`ees par les interf\`erences (milieux urbains), CoMP permet d'am\`eliorer le d\`ebit des utilisateurs en bord de cellule, leur fournissant une meilleure couverture et permettant une meilleure \`equit\`e dans le système. Les auteurs pr\`esentent une utilisation concrète de cette expression pour maximiser la capacit\`e de coupure, en optimisant le nombre de stations coop\`erantes et le d\`ebit de transmission.   } 

\RRmotcle{Distribution de capacit\`e, traitement multi-cellulaire, fading de Rayleigh} 
\RRabstract{In this paper we provide an analytical expression of the capacity outage probability for CoMP (Coordinated Multiple Point) multicell downlink transmission in Rayleigh fading. The theoretical derivation and simulation results match perfectly. Used in dense urban interference limited networks, CoMP is expected to improve cell-edge users' throughput and allow a better coverage and fairness. We demonstrate that our closed-form derivation can be helpful to maximise the capacity with outage and the goodput, by tuning appropriately the transmission data-rate and selecting optimal set of cooperating stations.} 
\RRkeyword{Multicell processing, capacity distribution, macro-diversity, Rayleigh fading.} 

\title{Outage Probability for Multi-Cell Processing under Rayleigh Fading}


\author{Virgile Garcia$^{1,2}$, Nikolai Lebedev$^{1,2,3}$, Jean-Marie Gorce$^{1,2}$\footnote{Contacts: virgile.garcia@insa-lyon.fr, lebedev@cpe.fr, jean-marie.gorce@insa-lyon.fr.}\\
	$^1$University of Lyon, INRIA \\
	$^{2}$INSA-Lyon, CITI, F-69621, France \\
	$^3$CPE Lyon,\ BP 2077, F-69616, France\\
	\textit{This paper has been submitted to IEEE Communication Letters}\\ \textit{and might be modified or unavailable without notice.}}
\date{\today}

\maketitle
\pagestyle{headings}
\begin{abstract}
In this paper we provide an analytical expression of the capacity outage probability for CoMP (Coordinated Multiple Point) multicell downlink transmission in Rayleigh fading. The theoretical derivation and simulation results match perfectly. Used in dense urban interference limited networks, CoMP is expected to improve cell-edge users' throughput and allow a better coverage and fairness. We demonstrate that our closed-form derivation can be helpful to maximise the capacity with outage and the goodput, by tuning appropriately the transmission data-rate and selecting optimal set of cooperating stations.
\end{abstract}


\section{Introduction}


Facing the problem of spectral resource limitation and continuous growth of user throughput, it is commonly recognized that the solution comes with smaller cells and dense spatial frequency reuse, possibly with the whole spectrum resource made available within each cell (i.e. Reuse 1). However, this increasing number of base-stations (BS) with shortened coverage range critically increases the inter-cell interference level, especially at cell boundaries.

The deployment of these small BS in urban areas imposes new difficulties for providers to plan their network, since they hardly control the location of the transmitters, their orientation or the propagation environment. This has motivated the development of the concept of Self-Optimized and Self-Organised Networks (SON)~\cite{3GPP-SON}, together with the necessity of an automated network planning and optimisation procedures with minimal human involvement.

Multi-cell processing, also called Coordinated Multiple Point (CoMP), is a very promising distributed multi-antennas technique that takes advantage of neighbour cell's antenna to perform a joint transmission and therefore increase users signal quality. This is expected to be part of next generation cellular networks standards such as LTE-A. 

Small cells networks in dense urban environments are mainly limited by interference. In these conditions, CoMP is expected to improve cell-edge users' throughput: a mobile located close to cell border would be able to decode a joint transmission from two or more BS within the same frequency band. That way, a better coverage is ensured with less users experiencing outage, and thus, a more intense reuse of the spectrum in dense areas is allowed. Several cooperation techniques (including CoMP) have been studied in~\cite{TseEURASIP08, shamai_enhancingcellular_2001, bjornson-globecom-2009, merouane-networkMIMO-2010}.


The main drawback of CoMP is the need for coordination between cells: a large amount of data representing users channel state information and algorithms related parameters are required, which could make it infeasible considering backhaul limitations. For evolved MIMO-like coordination, a perfect time synchronisation can be required to distributively generate beams or efficient precoding pattern~\cite{bjornson-globecom-2009}, which is a strong constraint. Looking for a practical solution, we focus on a simple Multi-Cell Processing that solely requires user mean SINR, and can be used with a less precise synchronisation than any beamforming-based MIMO. 

The main contribution of this paper is the analytic expression of the capacity outage probability in Rayleigh fading conditions for open-loop CoMP (Theorem~\ref{thm:outage_proba}). To the best of authors knowledge, such a derivation have not been pointed out in the literature. A derivation for the Single-Input Single-Output~(SISO) links was made in \cite{Kandukuri2002}. We derive this outage probability for distributed Multiple-Input Single-Output~(MISO) links, using the generalized chi-square random variable~\cite{Hammarwall-GeneralChi2}. As well as for the CoMP technique studied here, only the average received powers are required as input of this theoretical expression.

The next section presents the signal model. Afterwards, we derive the SINR outage probability, and show how this expression can be used to tune the transmission data-rate and set of cooperative stations to fulfil the outage probability requirements. Finally, a numerical example in small cells conditions is provided, looking at fixed outage probability and optimised ``goodput''.
%
\section{Signal model}
In LTE, physical resources are allocated by time-frequency blocks, using OFDMA. Under block-fading and the average received power made available at transmitters, the scheduling scheme consists of random selection of blocks to be allocated to the mobiles. For a given user, we note $\mathcal{N}$ the set of $N=\mathrm{card}(\mathcal{N})$ BSs that serve this user in a coordinated manner, thus creating a distributed MISO link. At a given mobile, on a given resource block, the received complex signal is:
\begin{align}
	y= \sum_{n\in \mathcal{N}}h_{n} \sqrt{p_{n} g_{n}} x_{n}
	+\sum_{k\notin \mathcal{N}}h_{k} \sqrt{p_{k} g_{k}} x_{k}+ z,
\end{align}
where $h_{n}$ are iid circular symmetric complex Gaussian Random Variables~(RV), representing fast-fading; $x_{n}$ are the transmitted symbols; $p_{n}$ the transmitted power; $g_{n}$ the path gain, including the path-loss from antenna $n$ to the user and a log-normal shadowing (slowly varying with respect to fading); $z$ is the thermal noise, a normal variable of standard deviation $\sigma_z$. Users are by default attached to the BS with the strongest signal averaged over fading $\arg\max_n (P_{n})$, where $P_{n}=p_{n}\ g_{n}$.

We assume that BSs are aware only of a partial CSI, consisting of the average received power fed back by the user, and not of the instantaneous channel gain, since it would require a tremendous feedback and instant sharing of information between all cells. This implies that evolved CoMP techniques such as Zero-Forcing, Beam-Forming or Dirty Paper Coding \cite{TseEURASIP08, bjornson-globecom-2009} are not suitable in this context. Perfect CSI knowledge and coherent detection are assumed at receiver side. The received SINR is then:
\begin{equation}
	\gamma(\mathcal{N})=
	\frac{\sum_{n\in \mathcal{N}} \vert h_{n}\vert ^2 P_{n}}
		 {\sum_{k\notin \mathcal{N}} \vert h_{k}\vert ^2 P_{k}+ \sigma_z^2}
	\label{eq:general_sinr}
\end{equation}
One of the challenges in CoMP is to determine the best number of cooperating stations since the improvement in SINR is balanced by the increased cost of resource due to cooperation. Each mobile can determine its own number of coordinating BSs, depending on the topology of the surrounding BSs.
\section{Outage in Rayleigh Fading}
\subsection{Outage probability}
Considering Rayleigh fading, $H_{n}=\vert h_{n}\vert ^2 P_{n}, \forall n$, is a chi-squared RV of order 2 with zero mean and a variance $P_{n}$. 

In \cite{Kandukuri2002}, authors derived $\mathbb{P}(SIR\leq\gamma_{th})$, the outage probability for a SISO link experiencing Rayleigh fading:
\begin{equation}
	\mathbb{P}(SIR\leq \gamma_{th})\!=\!\mathbb{P}\bigg(\frac{H_1}{\sum_{k=2}^N H_k}\leqslant\gamma_{th}\!\bigg)
	\!=\!1\!-\!\prod_{k=2}^N \frac{P_1}{P_1+\gamma_{th}P_k}
	\label{eq:boyd_equation}
\end{equation}
The thermal noise was neglected within the interference limited conditions, but can easily be taken into account.

In our case, the received SINR (\ref{eq:general_sinr}) is a non-trivial RV, composed by the fraction of sums of independent, non identically distributed chi-squared random variables.  
The pdf of the sum of generalized chi-square random variables~\cite{Hammarwall-GeneralChi2} is:
\begin{eqnarray}
	\mathbb{P}\bigg(\sum_{n\in \mathcal{N}}H_{n}=x\bigg) = \sum_{n\in \mathcal{N}} \frac{e^{-x/P_n}}{P_n \prod_{j\in \mathcal{N}, j\neq n}(1-\frac{P_j}{P_n})}
\end{eqnarray}
Note that mathematically, each received power must be different from another.
We can consider without loss of generality that this condition is satisfied, since in our case, transmitting antennas are not co-located.
\begin{thm}\label{thm:p_out}
The outage probability of a user with a set $\mathcal{N}$ of cooperating BSs and minimum required SINR $\gamma_{th}$ is:
\begin{align}
&P^{out}_{\mathcal{N}}(\gamma_{th}) =\notag\\
&1-\sum_{n\in \mathcal{N}}\bigg(e^{- \frac{\gamma_{th}\sigma_z^2}{P_{n}}} \!\!\!\!\!\!
\prod_{j\in \mathcal{N}, j\neq n}\frac{P_{n}}{P_{n}-P_{j}}
\prod_{k\notin{\mathcal{N}}} \frac{P_{n}}{P_{k} \gamma_{th}+P_{n}}
\bigg)
\label{eq:P_out}
\end{align}
Proof: see Appendix~\ref{sec:Proof}.
\label{thm:outage_proba}
\end{thm}
This CDF of SINR applies for a given OFDMA block. To obtain the capacity outage probability in case of several independently faded blocks allocated to a user, Theorem \ref{thm:outage_proba} can be extended, using the integration of convolutions of the PDF of the capacity.
\subsection{Spectral efficiency optimisation}
With the result given by (\ref{eq:P_out}), one can precisely calculate the exact outage capacity for a given BS' transmit powers scenario. A common way to evaluate capacity with outage~\cite[4.2.3]{WComGoldsmith} is to select an arbitrary outage probability (e.g. $1\%, 10\%$...) and look at the corresponding capacity. The outage probability enables us to evaluate the optimal \textit{goodput} $G$ (bit/s/Hz), which is a useful capacity metric for fading wireless link: the goodput is the successfully received capacity, taking into account the error probability of the transmission rate. For a given user with a set $\mathcal{N}$, the maximum goodput is obtained by optimising the transmission rate $R=\log_2(1+\gamma)$:
\begin{eqnarray}
	&G(\mathcal{N}) &= \max_{R}\ R\big(1-P^{out}_{\mathcal{N}}(2^R-1)\big)\nonumber\\
	&				&= \max_{\gamma}\ \log_2(1+\gamma)\big(1-P^{out}_{\mathcal{N}}(\gamma)\big)
	\label{eq:goodput}
\end{eqnarray}
When we consider a fixed outage probability $p_o$, $G(\mathcal{N})$ refers to the capacity $\log_2(1+\gamma_o)(1-p_o)$, for which $P^{out}_{\mathcal{N}}(\gamma_o)=p_o$.


To fairly compare performance of mobiles attached to one or several BSs, we need to take into account the cost of resource it induces: consider a cellular network in which the mobiles can cooperate with two BS. If we decide to attach the mobile to two BSs instead of one, this mobile will \textit{consume} twice the resource allocated to it: one by each of the cooperating BS. 
Thus, interested in spectral usage, we look at a per-BS spectral efficiency.
User's spectral efficiency $C$(bit/s/Hz/BS) is:
\begin{equation}
	C(\mathcal{N}) = \frac{G(\mathcal{N})}{\mathrm{card}(\mathcal{N})}
	\label{eq:spectral_eff}
\end{equation}


Taking into account the simple case where no power control is performed (as in 802.11g or for hardly manageable small cells), average received power are supposed known and fixed. To optimise user's spectral efficiency in that case, we then just have to select the best set of antennas $\mathcal{N}^*$, such that 
\begin{equation}
	\mathcal{N}^*=\arg\max_{\mathcal{N}} \frac{G(\mathcal{N})}{\mathrm{card}(\mathcal{N})}
\end{equation}
Note that, in general, for each user the set $\mathcal{N}^*$ and the number $N$ of cooperating BS is different; but this set is not necessary the one that maximise the goodput or the capacity with outage.

For a more evolved case with power control, theorem~\ref{thm:outage_proba} holds, as any received power values can be taken as input. A joint transmit power and multiple BSs association optimisation has to be made, but is out of the scope of this letter.

\section{Numerical results}
To illustrate a usage of (\ref{eq:P_out}), we numerically simulate received powers at mobiles with no power control applied. We use a 2D area with BSs uniformly distributed with a density of 100~BS/km$^2$~(which is similar to an hexagonal scheme with $124$~m inter-cell distance). The path loss model~\cite{3GPP25.996} used is $g_{dB}(d)=-(34.53+38\log_{10}(d)+\delta)$, where $\delta$ represents the log-normal shadowing. Up to $N=8$ cooperating BSs are considered here for practical reasons.

\begin{figure}
	\includegraphics[width=0.9\linewidth]{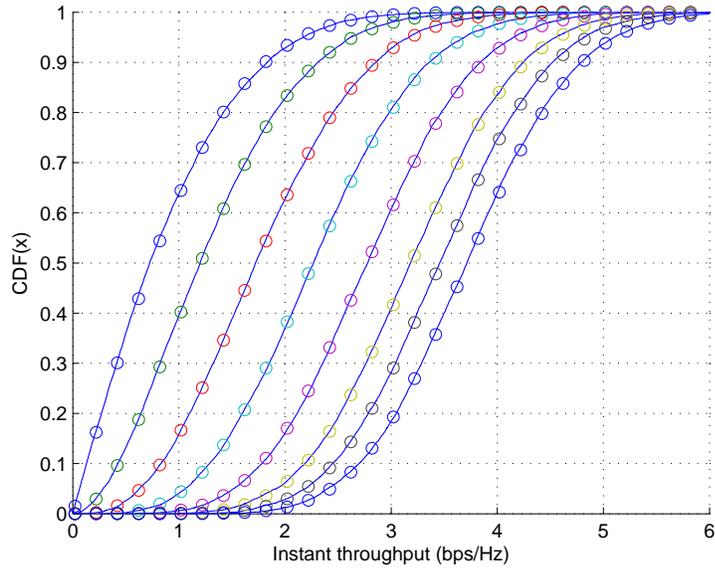}
	\caption{Comparison between empirical and theoretical CDF of a user capacity, for $N=\lbrace1,\cdots ,8\rbrace$ (from left to right).}
	\label{fig:fit}
\end{figure}

Fig.\ref{fig:fit} compares the analytical expression of outage capacity (\ref{eq:P_out}) to simulation obtained with a large number of realizations of each RV in (\ref{eq:general_sinr}), for arbitrary chosen user. This figure shows the perfect match of analytical and simulations results.

\begin{figure}
	\includegraphics[width=0.9\linewidth]{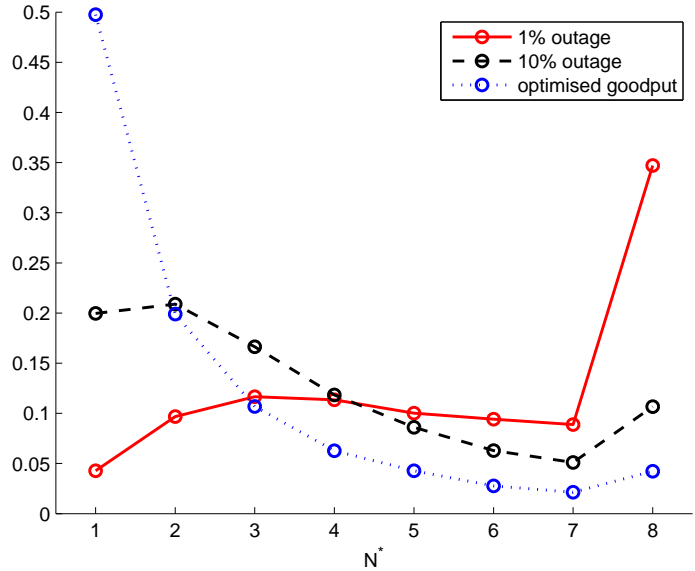}
	\caption{Fraction of users selecting $N^*$ cooperating stations, depending on the outage required.}
	\label{fig:proba_CoMP_nb_ant}
\end{figure}

Fig.\ref{fig:proba_CoMP_nb_ant} shows how users in small cells select their optimal number of BS. We can observe that users requiring a small outage probability (typically real-time applications such as voice or video streaming) select a higher number of cooperating BSs to maximize their outage spectral efficiency. Applications that are more tolerant to instant data losses (e.g. file transfers), benefit from fading diversity. The latter are expected to require the smaller number of cooperating BS (around $50\%$ of users keeping only one attachment BS).

\section{Conclusion}
In this paper we derive a theoretical expression for outage capacity in Rayleigh fading with a CoMP downlink transmission. The only available CSI-T are the average powers of cooperating BSs. This expression is used to determine the set of BSs that offer the maximum spectral efficiency among other combinations.


\section*{Acknowledgement}
This work has been carried out in the frame of the joint lab between INRIA and Alcatel-Lucent Bell Labs on ``Self Organizing Networks''.

{\section*{Proof of Theorem 1.}\label{sec:Proof}}
Let us start with the derivation of the complementary cumulative density function of generalised chi-square RV:
\begin{align}
	\mathbb{P}\bigg(\sum_{n\in \mathcal{N}} H_n> x\bigg)
	&= \sum_{n\in \mathcal{N}} \frac{e^{-x/P_n}}{\prod_{j\in \mathcal{N}, j\neq n}(1-\frac{P_j}{P_n})}
	\label{eq:CCDF}
\end{align}
Using (\ref{eq:general_sinr}) as SINR and the same reasoning as \cite{Kandukuri2002}'s derivation for SISO, we derive the success probability to reach the threshold rate $R_{th}~=~\log_2(1+\gamma_{th})$ for a distributed MISO link, every $H_n$ being independent:
\begin{align}
	&P^{out}_{\mathcal{N}}(\gamma_{th})=
	\mathbb{P}(\gamma(\mathcal{N})>\gamma_{th})=& \nonumber \\
	&\mathbb{P}\bigg(\sum_{n\in \mathcal{N}}H_n
			> \gamma_{th} \Big(\sum_{k\notin{\mathcal{N}}}H_k+\sigma_z^2\Big)\bigg)=& \nonumber \\
	&\int_0^{\infty}\!\!\!\!\!\!\! \ldots\!\! \int_0^{\infty} 
			\!\!\mathbb{P}\bigg(\sum_{n\in \mathcal{N}} H_n
			> \gamma_{th} \Big(\sum_{k \notin{\mathcal{N}}}t_k+\sigma_z^2\Big)\bigg)
			\!\prod_{k\notin{\mathcal{N}}} \frac{e^{-t_k/P_k}}{P_k} dt_k&  \nonumber
\end{align}
Using (\ref{eq:CCDF}), $P^{out}_{\mathcal{N}}(\gamma_{th})$ becomes:
\begin{align}			
	&\int_0^{\infty}\!\!\!\!\!\!\! \ldots\!\! \int_0^{\infty} 
			\!\sum_{n\in \mathcal{N}} \frac{e^{-\gamma_{th}\big(\sum_{k \notin{\mathcal{N}}}t_k+\sigma_z^2\big)/P_n}}
			{\prod_{j\in \mathcal{N}, j\neq n}\big(1-\frac{P_j}{P_n}\big)}
			\prod_{k\notin{\mathcal{N}}} \frac{e^{-t_k/P_k}}{P_k} dt_k=& 	\nonumber \\
	&\sum_{n\in \mathcal{N}} \frac{e^{-\gamma_{th}\sigma_z^2/P_n}}
			{\prod_{j\in \mathcal{N}, j\neq n}(1-\frac{P_j}{P_n})} 
			\prod_{k\notin{\mathcal{N}}}
			\int_{t_k=0}^\infty \frac{e^{-t_k (\gamma_{th} /P_n+1/P_k)}}{P_k} dt_k=& \nonumber\\
	&\sum_{n\in \mathcal{N}} 
	e^{\frac{-\gamma_{th}\sigma_z^2}{P_n}} \prod_{j\in \mathcal{N}, j\neq n}\frac{1}{1-\frac{P_j}{P_n}}
	\prod_{k\notin{\mathcal{N}}} \frac{1}{(1+\gamma_{th}\frac{P_k}{P_n})}&
	\label{eq:outage_calcul}
\end{align}
Rearranging the $P$ factors in (\ref{eq:outage_calcul}) gives (\ref{eq:P_out}).




	\bibliographystyle{unsrt}
	\bibliography{biblio}

%
%
%
%
%




\end{document}